\documentclass[useAMS]{mn2e}
\usepackage{graphicx}

\title[Photometry from online
Digitized Sky Survey Plates]{Photometry from online Digitized Sky
Survey Plates}
\author[A. Bacher, S. Kimeswenger and P. Teutsch]
{A. Bacher, S. Kimeswenger and P. Teutsch
\\
Institut f\"ur Astrophysik der
Leopold--Franzens--Universit\"at Innsbruck, Technikerstr. 25,
A-6020 Innsbruck, Austria}
\begin{document}

\date{Accepted . Received}

\pagerange{\pageref{firstpage}--\pageref{lastpage}} \pubyear{2005}

\maketitle

\label{firstpage}

\begin{abstract}
Online Digital Sky Survey (DSS) material is often used to obtain
information on newly discovered variable stars for older epochs
(e.g. Nova progenitors, flare stars, \dots). We present here the
results of an investigation of photometry on online digital sky
survey material in small fields calibrated by CCD sequences. We
compared different source extraction mechanisms and found, that
even down near to the sensitivity limit, despite the H-compression
used for the online material, photometry with an accuracy better
than 0\fm1 rms is possible on DSS-II. Our investigation shows that
the accuracy depends strongly on the source extraction method. The
SuperCOSMOS scans, although retrieved with an higher spatial
resolution, do not give us better results. The methods and
parameters presented here, allow the user to obtain good plate
photometry in small fields down to the Schmidt plate survey limits
with a few bright CCD calibrators, which may be calibrated with
amateur size telescopes. Especially for the events mentioned
above, new field photometry for calibration purposes mostly
exists, but the progenitors were not measured photometrically
before. Also the follow up whether stellar concentrations are
newly detected clusters or similar work may be done without using
mid size telescopes. The calibration presented here is a "local"
one for small fields. We show that this method presented here
gives higher accuracies than "global" calibrations of surveys
(e.g. GSC-II, SuperCOSMOS and USNO-B).
\end{abstract}


%
\sloppy
\section{Introduction}
The photographic sky surveys produced by large Schmidt telescopes
have proved to be among the most useful and enduring weapons in
the astronomical armory. With the start of the Digitized Sky
Survey (hereafter DSS; Lasker \& McLean 1994) new powerful tools
have been established. To provide convenient access to these data,
the images have been compressed using a technique based on the
\mbox{H-transform} to reduce the data volume. Although the
technique is lossy, it is adaptive so that it preserves the signal
very well. They typically compress the data by a factor of 7, but
much higher compression ratios are possible. The southern surveys
are covered by the SuperCOSMOS Sky Survey (hereafter SSS; Hambly
et al. 2001a), too. This survey provides us with images at even
higher spatial resolution. While there are extensive studies on
the photometric calibration of whole plates or even surveys on
original uncompressed data (e.g. Reid \& Gilmore 1982, Lasker et
al. 1990, Russell et al. 1990, Morrison et al. 1997, Monet et al.
2003 = \mbox{USNO-B}), the possibilities of local small field
calibration on compressed online available material were not
detailed studied for stellar photometry. Galaxy surveys use this
material more often and thus much better studies are available
there (e.g. Kron 1980, Spagna et al. 1996, Hambly et al. 2001b).
Although there is often the need to look up the older epoch data,
especially for eruptive variable stars, only limited effort on the
source extraction and calibration technique for stars was done
(e.g. King et al. 1981, Humphreys et al. 1991, H\"ortnagl et al.
1992, Kimeswenger \& Weinberger 2001, Andersen \& Kimeswenger
2001, Kimeswenger et al. 2002a, 2002b, 2003). The goal of this
study here is to show how to optimize source extraction parameters
in small fields around interesting sources. This allows a user to
derive a high quality local calibration of online digitized
plates, having only a few bright calibration sources in the field.
Such calibrators can be obtained easily by CCD cameras at amateur
size telescopes or training equipment of universities (Kimeswenger
2001, Bacher et al. 2001, Lederle \& Kimeswenger 2003).

\section{Data}
The photometric data was obtained from the online servers at the
STScI ({\small\sl
http://archive.stsci.edu/cgi-bin/dss\_plate\_finder}), ESO
({\small\sl http://archive.eso.org/dss/dss}) and ROE ({\small\sl
http://www-wfau.roe.ac.uk/sss/pixel.html}). The characteristics of
these surveys are summarized in Table~\ref{tab_color_eq}.

\begin{table*}
\caption{Parameters and colour equations for sky survey plates used
here. Where several colour equations are given, always the first
one was used here (see text). Some SuperCOSMOS scans are only
partly available yet} \label{tab_color_eq}

\begin{tabular}{llcccll}
survey & plate/filter & band & {\scriptsize DSS} scan & {\scriptsize SuperCOSMOS} & colour equation & reference \\
       &              &      & resolution &  resolution &        & \\
\hline
Pal-QV No & IIaD+W12     & V            &  1\farcs7       &           &  \\
SERC-J/EJ    & IIIaJ+GG395  & B$_{\rm J}$  &  1\farcs7       & 0\farcs67 &  B$_{\rm J}$ = B - 0.28 (B-V)             & Blair \& Gilmore \cite{blair82} \\
             &              &              &            &           &  B$_{\rm J}$ = B - 0.20 (B-V)             & King et al. \cite{king81} \\
             &              &              &            &           &  B$_{\rm J}$ = B - 0.23 (B-V)             & Kron \cite{kron80} \\
POSS II J    & IIIaJ+GG385  & B$_{\rm J}$  &  1\farcs0       &           &  B$_{\rm J}$ = B - 0.28 (B-V)             & Reid et al. \cite{reid91} \\
POSS I E     & 103aE        & E            &  1\farcs7       & 0\farcs67      &  E = R$_{\rm c}$                          & Spagna et al. \cite{spagna96} \\
SERC-ER/R    & IIIaF+OG590  & r = R$_{\rm 59}$ &  1\farcs0   & 0\farcs67      &  r = R$_{\rm c}$ + 0.04 (V - R$_{\rm c}$) & H\"ortnagl et al. \cite{HKW92} \\
POSS II R    & IIIaF+RG610  & r = R$_{\rm 61}$ &  1\farcs0   &           &  used same as SERC-R                      &  \\

\hline
\end{tabular}
\end{table*}

As photometric sequences those of Henden \cite{aah02} were
chosen. These widely used sequences were compared and tested with
own measurements from the Innsbruck 60cm telescope (Kimeswenger et
al. 2002a, Lederle \& Kimeswenger 2003) around V838\
Mon and CI\ Aql and with those by K.S. from ESO NTT around V4332\
Sgr.

Colour equations for plate/filter combinations were obtained by
various authors in different manner. While King et al.
\cite{king81} have arbitrarily chosen a relation based on visual
estimate from the passbands, H\"ortnagl et al. \cite{HKW92} and
Henden \cite{aah03} for use in USNO~B (Monet et al. 2003) used
spectrophotometric catalogues to fold with published filter +
detector response curves. Kron \cite{kron80}, Blair \& Gilmore
\cite{blair82}, Reid et al. \cite{reid91}, Humphreys et al.
\cite{mag_dia} and Spagna et al. \cite{spagna96} used
photoelectrical or CCD photometry of stars in standard bands to
derive the equations. Where available we used the latter ones
here. We kept the native magnitude system of the plates and
converted the CCD magnitudes to this system since the POSS-II and
the southern plates for a given field were taken at different
epochs, so variables could get really weird colours, along with
some other factors.

\section{Calibration Method}

\begin{figure}
\centering
\includegraphics[width=7.8cm]{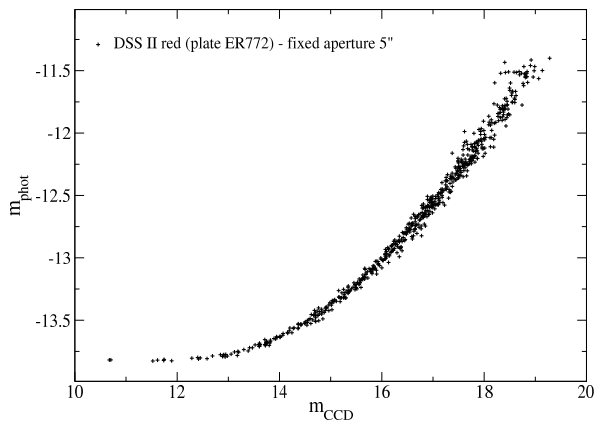}
\caption{A sample of photographic (DSS-II red) fixed aperture
(5\arcsec) magnitudes vs. CCD magnitudes. While the faint end
gives us a linear relation, the core saturation of the bright
stars lead to an asymptotic convergence to a constant value at the
bright end. } \label{fig1}
\end{figure}

The source extraction was performed using SExtractor V2 (Bertin \&
Arnouts 1996). We used aperture photometry with fixed sized
apertures and Kron first moment adaptive radius apertures (Kron
1980) as implemented to SExtractor. Due to the core saturation of
bright stars we expect a curve reaching asymptotically a constant
value at the bright end and being quasi linear at the faint end
(see Figure~\ref{fig1}).

Such a curve was "invented" for the calibration of photographic
densities by Moffat \cite{Moffat}. He investigated in detail the
calibration curves and the image growth of the stars on Hamburg
Schmidt plates. We here use the equation of the same form

\begin{equation}
m_{\rm phot} = a_0 \times \log\left({10^{a_1 \times (m_{\rm CCD} -
a_2)} + 1}\right) + a_3
\end{equation}

\noindent where $m_{\rm phot}$ is the magnitude coming directly
from the source extraction on the DSS or SuperCOSMOS plate,
$m_{CCD}$ is the CCD magnitude corrected to the colour system of
the plate and $a_i;\,\, i \in [0,3]$ are free fit parameters. As
we measure in case of stars an integrated value

\begin{equation}
\int_0^{R_{\rm aperture}}{f(T(r))\, r \, {\rm d}r} -
\int_0^{R_{\rm aperture}}{f(S(r))\, r \, {\rm d}r}
\end{equation}

\noindent were $T(r)$ is the really transmitted light in the
measurement machine for the star { and $S(r)$ that of the sky
background. In the software we use the sky background is
calculated as 2D polynomial in the region around the stellar
aperture after exclusion of contamination by other sources.
Details of this procedure can be found in Bertin \& Arnaults
(1996).} This transmission is a function of seeing - and as shown
by Moffat (1969) and Kimeswenger (1990) - strongly changes due to
the wavelength dependent scattering within the plate. As shown
already by Moffat and more in detail by Irwin \& Hall (1983)
simple fits of this growth function with e.g. Gaussians (as
working fine for direct imaging CCDs) do not work well.

 Assuming $a_1 \rightarrow
1$ leads to the linearized "Baker densities". This works fine in
case of some emulsion and wavelength domains (Hoffman et al.
1998). This fact was widely used to "linearize" the calibration
curves (e.g. the COSMAG values in the SuperCOSMOS catalogue) and
to correct for the remaining deviation by semi-empirical functions
(Hambly et al. 2001a). In fact this leads to "global" parameters
for the liberalization and free parameters fitting deviations.
This is very suitable for global calibrations but not needed for
the intended local calibration here.

But our main uncertainty in case of the public DSS data in fact is
the unknown conversion function $f$ in the integration (Equ. 2)
above. We measured the value of the unexposed plate (thereafter
called chemical fog) and found strong variations from plate to
plate. The machines were not calibrated in the same way all the
time when scanning DSS2. Thus a general clear correlation of our
fit parameters with the theoretical values by Moffat (1969) is not
possible. This seems to be a major difference to the SuperCOSMOS
scans, were all plates seemed to be scanned and calibrated in the
same way. Therefore there is no way around a fitting of the
parameters to obtain a proper solution.

\begin{figure} \centerline{
\includegraphics[width=6.8cm]{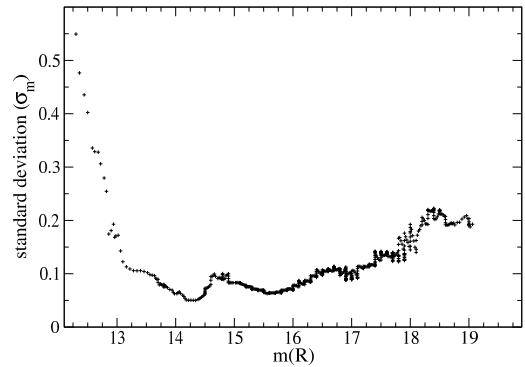}}
\caption{The rms as function of magnitude for plate ER772 in the
region around the variable star V838 Mon. The error mostly lies at
or below 0\fm1. The slow increase towards the fainter end is
partly denoted to the accuracy of the photoelectric sequence of
Henden (2002). At the bright end the error increases rapidly due
to the core saturation. { The smaller "jump" (at 14\fm5) is due to
a "cluster" of slightly blended stars around the target V838 Mon,
having all about the same apparent magnitude. They also might have
inaccurate CCD magnitudes, as the calibration sequence was taken
during a bright outburst of the Target.}} \label{sigma}
\bigskip
\end{figure}

While $a_3$ gives an estimate for the core saturation value, which
strongly depends on the aperture size, $a_0 \times a_1$ represents
the inclination of the linear section, $a_1$ also influences the
curvature around the knee and $a_2$ gives more or less the start
of the linear section. An individual fit $m_{\rm phot}(m_{\rm
CCD}; a_0, a_1, a_2, a_3)$ for each CCD sequence and each aperture
was derived. The correlation coefficient for all fits was always
$R^2
> 0.995$. We derived the standard deviation of the fitted values
$\sigma_{\rm f}$ in bins of $\Delta m = 0\fm1$ for $m
> 12\fm0$. It is nearly negligible for bright sources and
increases steadily towards the exposure limit. Near the limit the
error will originate partly also from the CCD sequences (see
Figure~\ref{sigma}).

The reverse function of (1) then has to be used to derive the
$m_{\rm CCD}$ of targets
\begin{equation}
m_{\rm CCD} = {1 \over a_1} \times \log\left({10^{(m_{\rm phot} -
a_3)\over a_0} - 1}\right) + a_2
\end{equation}
To derive an error estimate for the resulting magnitudes
$\sigma_{\rm m}$ for the reverse function we used the derivative
\begin{equation}
\sigma_{\rm m} = \sigma_{\rm f} \left({{\rm d}m_{\rm phot} \over
{\rm d} m_{\rm CCD}}\right)^{-1} = {\sigma_{\rm f} \over a_0
a_1}\left({1 + 10^{-a_1 (m_{\rm CCD} - a_2)} }\right)
\end{equation}

This clearly leads to a very high effective error at the bright
end of the calibration where the saturation effects the data.
There the isophotal magnitudes are definitely better. But as we
have shown in the introduction we are not interested in this part
of the calibration as it can be reached better by other means. In
Figure~\ref{sigma} the rms residuals for a fit are shown. For most
of the brightness region of interest the error is about 0\fm1 or
even below. Finally we tested whether the fits, respectively the
residua are a function of the stellar colour. This test gives us
information on the quality of the colour equations used.

\begin{figure} \centerline{
\includegraphics[width=7.5cm]{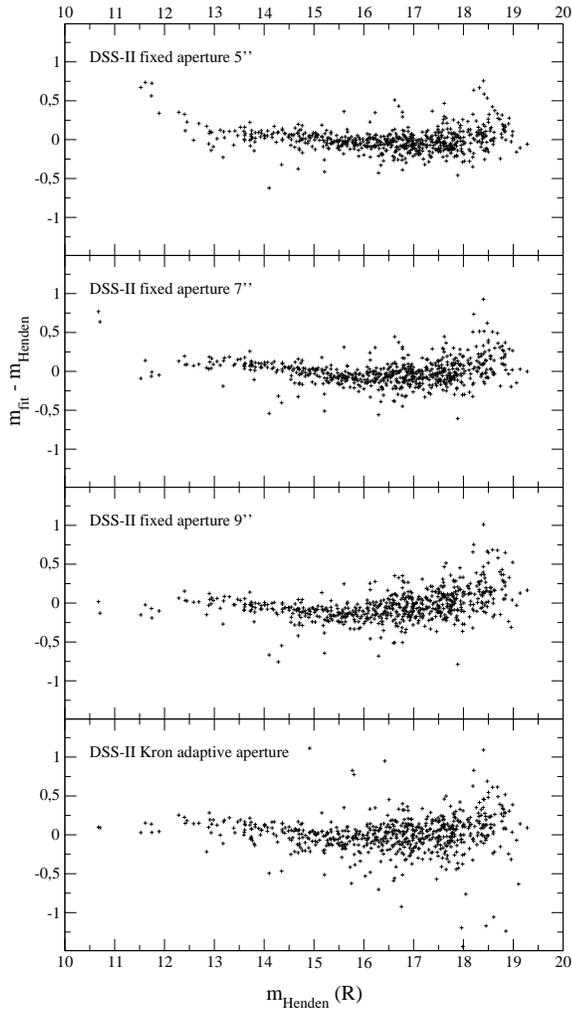}}
\caption{The deviation of fits with different apertures in the
15\arcmin$\times$15\arcmin field around V838~Mon.}
\label{deviation}
\end{figure}

\begin{figure}
\centering
\includegraphics[width=7.5cm]{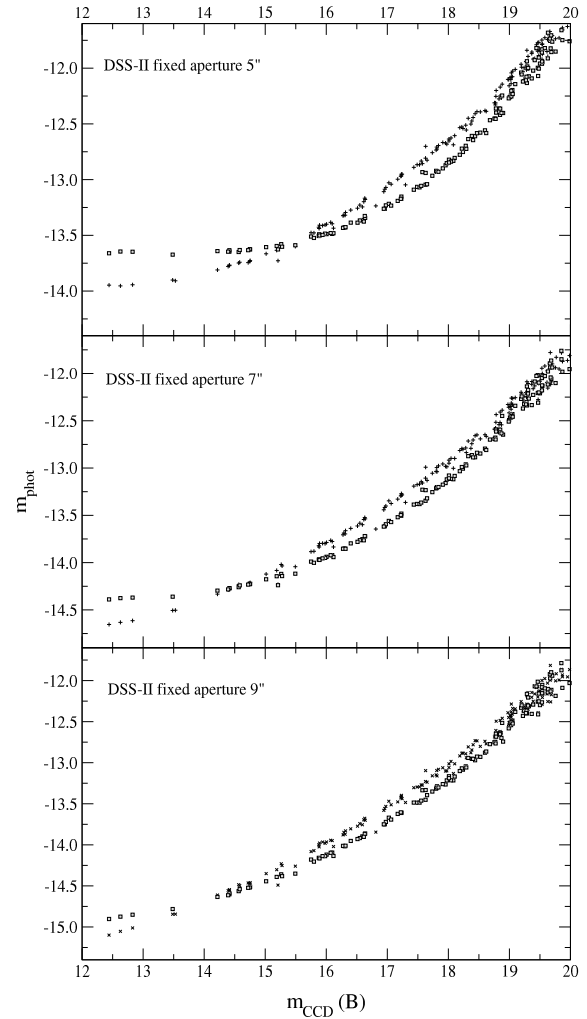}
\caption{The field around the variable star AM Her on the plates
XJ229 and XJ230. The overlapping region allows us to study the
different behaviour of the plates individually.} \label{fits_dss2}
\end{figure}

The smaller the aperture we used, the smaller the error was
(Figure~\ref{deviation}) at higher magnitudes. On the other hand
the saturation goes deeper. Often adaptive Kron apertures (Kron
1980) are used to avoid saturation. This method is also
implemented in SExtractor. Comparing our result to the accuracy
shows that the rms is larger at all wavelengths. Crowding seems to
confuse too often the adaptive apertures. Kron adaptive apertures,
using two or more threshold levels, are indeed a kind of profile
fitting. As already pointed out by Irwin \& Hall (1983), profile
fitting leads to the highest errors. They find best results for
isophotal photometry, while we find in our crowded fields in the
galactic plane best results by using small apertures.

The solutions for the individual scans strongly differ from plate
to plate. There is no general solution for the parameters. As
shown in Figure~\ref{fits_dss2} the plates are exposed and scanned
so differently that neither the saturation ($a_3$) not the
inclination at the higher magnitudes ($a_0 \times a_1$) nor the
curvature ($a_1$) or the position of the knee ($a_2$) is fixed
somehow. Normally one does not have deep CCD sequences as we used
here for the base study. Thus we tested the robustness of the fits
with a small number of bright CCD calibrators. But in fact we were
searching for an extrapolation if only a few bright CCD
calibrators are available. We thus tried to derive some of the
parameters by basic measurement on the plate itself. For this
purpose we obtained the value of the unexposed plate $I_{F}$ at
the plate edge (often called chemical fog), the value of the sky
background $I_{S}$ and \mbox{finally} that of the overexposed
plate $I_{O}$ in the center of a bright star nearby. { All those
measurements are used "as is" in the published files. These units
are the original natural analog/digital converter units folded by
the (to the user unknown) function $f$ (see equation 2).} {
Looking to equation (1) one can see that $a_3$ gives the magnitude
of a star were the overexposed core is just as big as the (fixed
size) aperture. Inserting the saturation and the sky level to
equation (1) this leads} together with the aperture $d$ (in units
of pixels) to:

\begin{equation}
a_3 = -2.5 \log\left[{{\pi d^2\over
4}\,\left({I_{O}-I_{S}}\right)}\right]
\end{equation}

\noindent { In our tests we found a strict correlation of the
chemical fog and the sky level with the parameter $a_2$.} The
dynamic range from the chemical fog to the { sky level}
($I_{S}-I_{F}$) gives us

\begin{equation}
m_{\rm phot}(a_2) = -2.5 \log\left[{{\pi d^2\over
4}\,\left({I_{S}-I_{F}}\right)}\right]
\end{equation}

\noindent The CCD magnitude of a star with $m_{\rm phot}(a_2)$ (or the mean of
a set of stars within $m_{\rm phot}(a_2) \pm 0\fm05$) is equal to
$a_2$. This gives a limit for the deepness of the CCD sequence
needed for the calibration.

\noindent Finally inserting the CCD magnitude of the curvature
point $a_2$ into equation (1) we get with the two equations (5)
and (6) above a solution for $a_0$
\begin{equation}
a_0 = -{2.5 \over \log(2)}\log\left[{{I_{S}-I_{F}}\over
{I_{O}-I_{S}}}\right]
\end{equation}
This parameter $a_0$ is independent from the aperture, as it
describes the inclination of the curve at stars much smaller than
the aperture. As we now were able to fix 3 (out of 4) parameters,
the fit of the remaining parameter $a_1$ was robust against
reducing the number and the limiting magnitude of the calibrators.
\begin{figure}
\centering
\includegraphics[width=8.3cm]{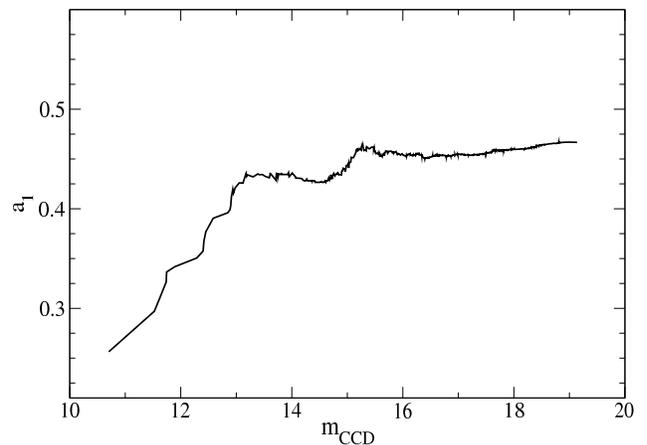}
\caption{The evolution of the fits of parameter $a_1$ as function
of the limiting magnitude. It converges to the optimum at
magnitudes around $a_2$ which was derived independently before as
described above.} \label{fits_a1}
\end{figure}

\noindent Figure~\ref{fits_a1} shows that the limiting magnitude
for the CCD sequence has to be deep enough to reach the value of
$a_2$ derived above. Deeper sequences do not improve the quality
of the parameter anymore. Thus a real extrapolation from bright
CCD sequences to faint stars is possible (see Figure 6).

\begin{figure}
\centering
\includegraphics[width=8.3cm]{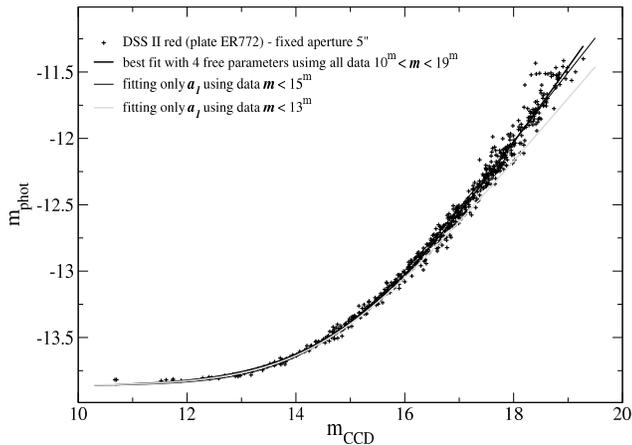}
\caption{The real 4 parameter fit is nearly identical with the fit
fixing 3 parameters and using the sequence up to $m_{\rm CCD} =
15\fm0$. Even the fit with very bright calibrators only, although
systematically wrong at the faint end, does not deviate more than
about 0\fm5 - the rms found often for globally calibrated
astrometric surveys like e.g. USNO~B.} \label{fits_accuracy}
\end{figure}

\section{The H-compression}

\begin{figure} \centerline{
\includegraphics[width=8.2cm]{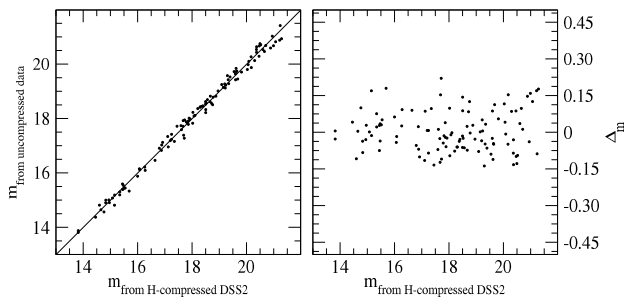}}
\caption{The photometry achieved on uncompressed { (SuperCOSMOS
plate rebinned to DSS resolution)} vs. DSS-II H-compressed images
around the suspected variable star V471\,Per using the method
described in section 3. There is obviously neither a systematic
effect nor any degrading of the photometric quality.}
\label{h-comp}
\end{figure}

The DSS data is compressed { by the use of a "lossy" algorithm}.
The H-compression (White et al. 1992) uses typical factors of 5 in
crowded fields and up to 10 at high galactic latitudes (Lasker et
al. 1996). We investigated the influence of the H-compression in
one field around the suspected symbiotic variable star V471\,Per.
{ For comparison the plate from the SuperCOSMOS data archive,
scanned at much higher resolution of 0\farcs67, was median
filtered 3$\times$3 pixels ($\equiv$ 1\farcs98$\times$1\farcs98)
and then rebinned to the DSS pixel size of 1\farcs01.} A CCD
calibration sequence is given by Henden \& Munari \cite{hemu01}.
The result in Figure~\ref{h-comp} shows that there is no
degeneration of the photometry even down to the plate limit. The
rms of the regression is below 0\fm09 at $m < 20\fm0$ and
increases to 0\fm14 at $20 \le m < 21\fm5$.

\section{Comparing to survey calibrations}
\begin{figure} \centerline{
\includegraphics[width=8.2cm]{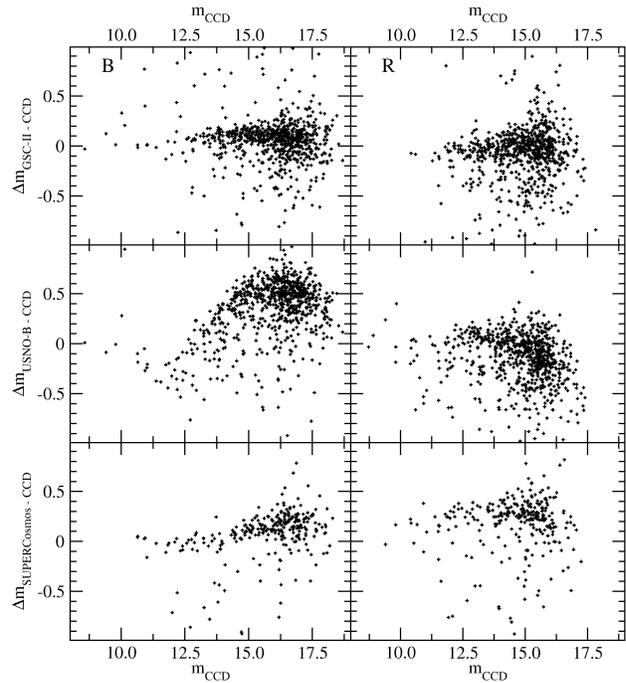}}
\caption{The deviation of the magnitudes given in the USNO-B
catalog (Monet et al. 2003), the GSC 2.2 (STScI, 2001), and the
SSS catalog (Hambly et al. 2001a) from those of Henden (2002) in
the 15\arcmin$\times$15\arcmin field around AM Her.}
\label{comp_ac_her}
\end{figure}

\begin{figure} \centerline{
\includegraphics[width=8.2cm]{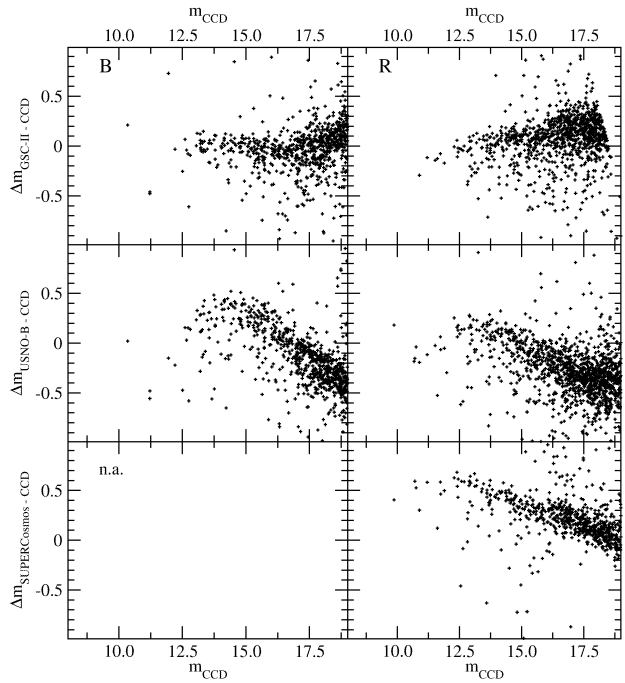}}
\caption{The deviation of the magnitudes given in the USNO-B
catalog (Monet et al. 2003), the GSC 2.2 (STScI, 2001), and the
SSS catalog (Hambly et al. 2001a) from those of Henden (2002) in
the 15\arcmin$\times$15\arcmin field around V838~Mon.}
\label{comp_v838}
\end{figure}

Recently a set of globally calibrated surveys was released. While
the GSC-II ({\small\sl
http://www-gsss.stsci.edu/gsc/gsc2/GSC2home.htm}) is based on the
same scans (but before H-compression) we used for our
investigation, the USNO-B catalog (Monet et al. 2003) and the SSS
catalog (Hambly et al. 2001a) used data from different scanning
machines. The USNO-B only uses an 8-bit linear converter for the
transmitted light. This undersamples the greyscale variations in
the centre of the stellar images. As already pointed out by Monet
et al. (2003), the photometry never was a main goal of this
survey. The SSS uses scans with a linear 15-bit greyscale from a
CCD. The data is pre-calibrated using an average slope for the so
called linear part of the emulsion of $\gamma = 2$. Then a complex
post-processing including CCD calibrators and colors is started
(Hambly et al. 2001a). This is essential for wide area statistical
studies, but not needed in our local calibration of individual
sources. The comparison of the Henden (2002) CCD sequences with
these catalogues for two of our fields is shown in
Figs.~\ref{comp_ac_her} \& \ref{comp_v838}.

The GSC2.2 do show a much higher noise but (nearly) no systematic
effects. The calibration is not well documented in the literature
neither at the survey homepage. Manual inspection of the sources
in our fields leads us to the conclusion that at least a part of
this error is originating from improper deblending of faint
sources. The typically higher crowding on the red survey plates
thus causes a higher error there.

The USNO-B photometry is based on extrapolation of bright TYCHO
stars and a few faint calibrators on some plates. Finally the
remaining plates were adjusted using the overlaps. We find
generally very high ($\ge 0\fm5$) systematic deviations for all
our fields.

The SSS uses a grid of faint CCD calibrators fields to define an
average gradient of the isophotal magnitudes vs. CCD magnitudes.
These were used to extrapolate towards the plate limit the
individual field-by-field calibration curves derived from bright
standards. Post-processing by a complex color classification and
the plate overlaps were used to do fine-tuning (Hambly et al
2001a). The comparison with our results in the small fields leads
to a much lower scatter than the surveys mentioned before. But
systematic effects are sometimes evident. This is already pointed
out by Hambly et al.: {\sl "The isophotal scale is non–linear at
the level of $>$ 0.5 mag; furthermore, such non–linearities are
not repeatable from field to field."}

\begin{figure} \centerline{
\includegraphics[width=3.9cm]{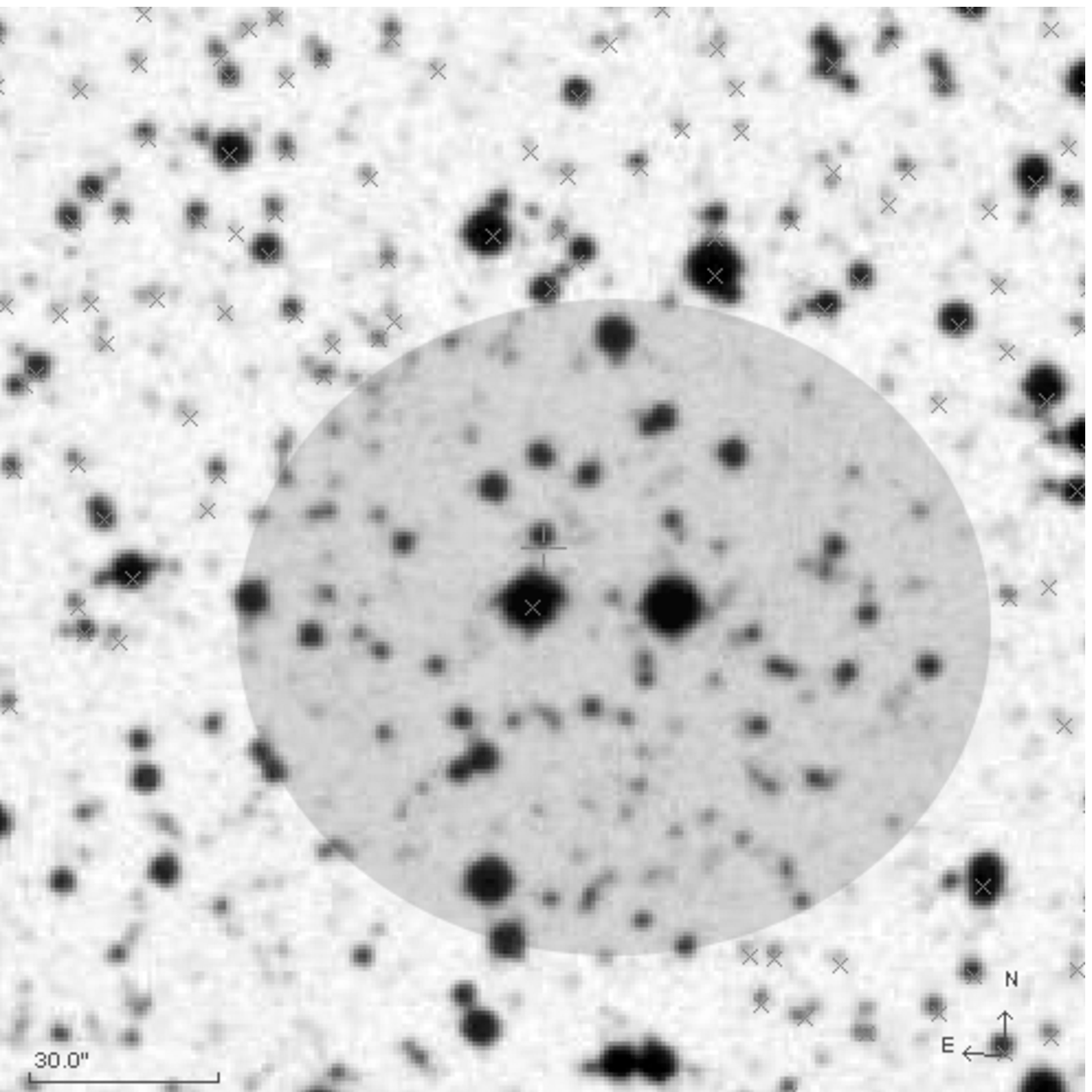}\phantom{XX}
\includegraphics[width=3.9cm]{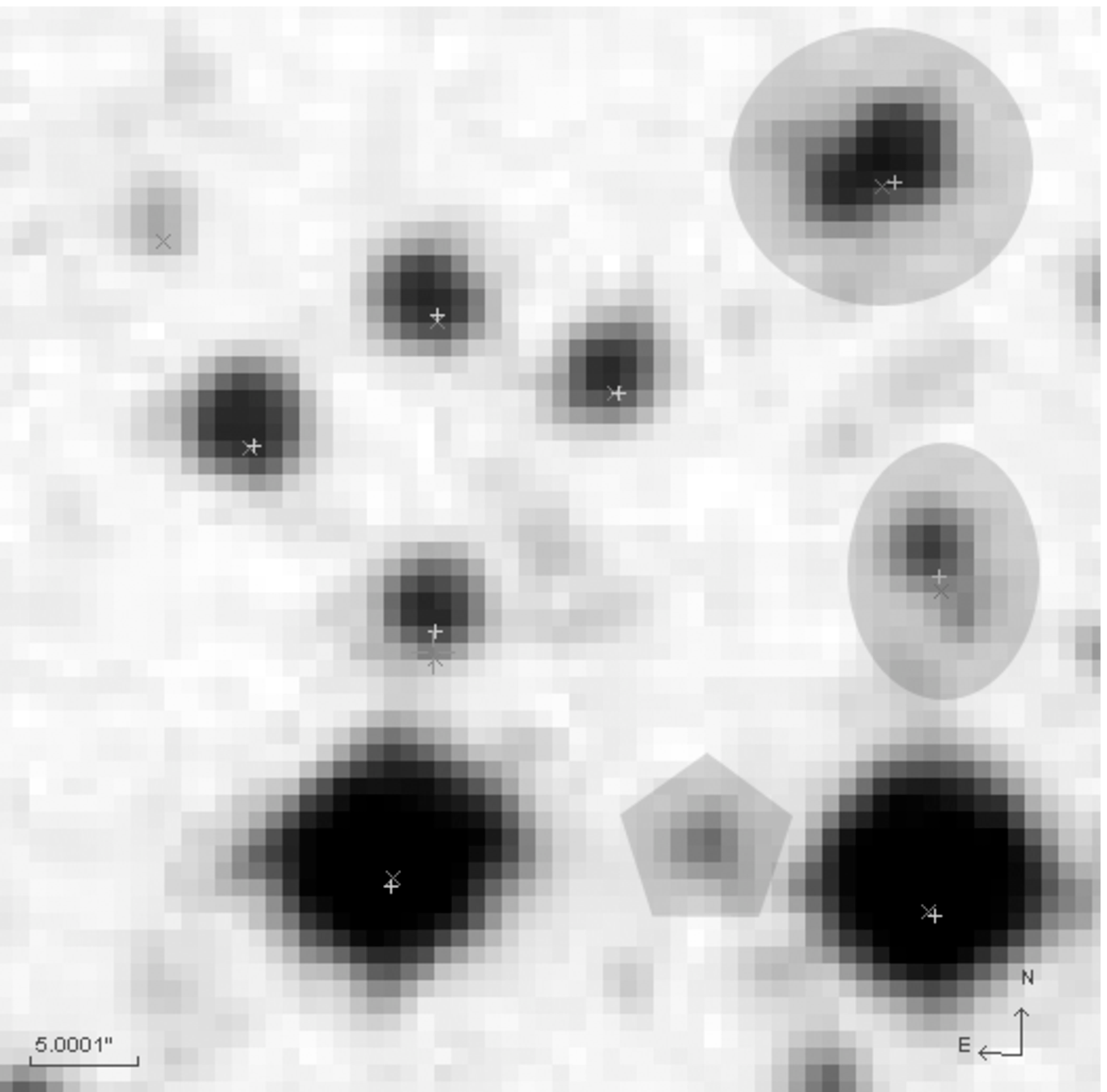}}
\caption{Left: The SSS avoids in the 'full-quality' catalogue the
region around bright stars (shaded area). This causes targets like
here V4332 Sgr being missed; Right: SSS and GSC-II do not deblend
targets, which are properly done by the neuronal network software
SExtractor (elliptical areas). Also faint stars having a good S/N
ratio nearby bright stars are ignored (polygonal area).}
\label{v4332_image}
\end{figure}

In a recent investigation of a field around the eruptive variable
V4332 Sgr (Kimeswenger, 2005) another effect was crucial. The SSS,
using the selection flags for high quality (= default), misses
targets near bright stars (Figure \ref{v4332_image}). Also
deblending, which works fine in SExtractor V2 (Bertin \& Arnouts
1996), is not implemented to GSC-II and SSS properly.

\begin{figure} \centerline{
\includegraphics[width=8.3cm]{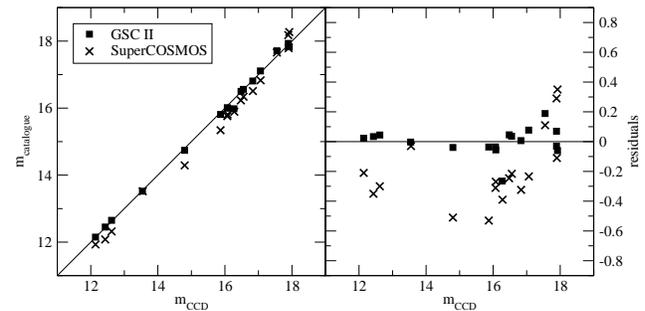}}
\caption{The GSC-II and the SSS red magnitudes in a field around
the eruptive variable V4332 Sgr (Kimeswenger, 2005).}
\label{v4332_calib}
\end{figure}

The calibration of the stars around V4332 Sgr by using ESO NTT CDD
images (limiting mag $\approx$ 24\fm0) obtained by SK gives, after
manual cleaning blends and all stars in the vicinity of 10" around
bright stars a very good result for GSC-II. Again the SSS has
strong systematic deviations (Figure \ref{v4332_calib}).

\section{Alternative calibration - CCD vs. GSC-II}
\begin{figure} \centerline{
\includegraphics[width=8.3cm]{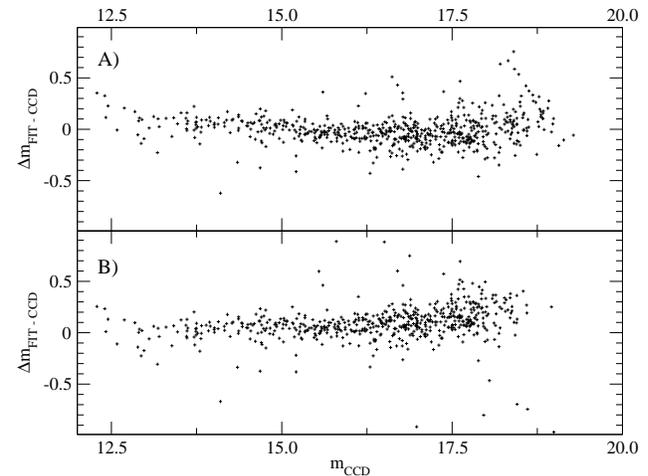}}
\caption{Calibration of the field around V838 Mon using the CCD
sequence (upper panel) and using GSC-II (lower panel). }
\label{gsc_calib}
\end{figure}

As shown in the previous section the GSC-II, although having a
higher rms error, shows nearly no systematic effects. Thus we
tested a calibration by use of the GSC-II instead of the CCD
sequence. This might be of interest in regions were no CCD
calibrations are available or for a first quick result - e.g.
especially after erruptive nova and nova like events. In
Figure~\ref{gsc_calib} the results of a calibration using a CCD
sequence and alternatively using only the GSC-II to derive the
$a_i$ parameters are shown. The results are promising. The
systematic effects are significantly below the noise. The accuracy
of such a local "post-processing" of the GSC-II improves the
photometry significantly.

\section{Conclusions}
The method we present here is well suited to calibrate locally
even H-compressed digital sky survey data by just using a few
bright calibrators. These calibrators easily can be achieved by
using small amateur telescopes. The accuracy is about 0\fm1 for a
wide range. Thus also colour-colour diagrams are possible. This
allows to do photometry of progenitors for eruptive events like
novae as well as HR diagrams when searching for new candidates for
stellar clusters. The importance for variable and eruptive star
studies was pointed out recently by Munari et al. \cite{mun},
Kimeswenger et al. \cite{v838} and Kimeswenger \& Lechner
\cite{nova03}. Goranskij et al. \cite{g04} shows the strong
implications on physical models of such kind of DSS usage. As
pointed out by Kimeswenger \cite{v4332} several different epochs
by using e.g. overlaps, not included all to the surveys, give us
important information for such variable objects. { Here we do not
focus on light curves of variables on a set of (similar) plates
with respect to a reference plate. In such a case photometric
differential work is usually the best choice. Comparing single
epoch data or data on different bands (e.g. POSS-I and POSS-II
blue are different) with new CCD data the progenitors of eruptive
variables indeed need absolute calibration.} But also the
investigations of small dark clouds by St\"uwe \cite{jogi}, who
used statistical calibrations with a BS model (Bahcall \& Soneira
1984) and complementary galactic structure work like that of Curry
\& McKee \cite{c00} and of Castellani et al. \cite{c01} might be
improved by this kind of calibrations presented here.\\
The local calibration presented here is clearly not overcoming
global surveys (e.g. Hambly et al. 2001a or Monet et al. 2003) for
large statistical samples, but due to the higher accuracy
(restricted to small fields) a perfect access for studies of
individual targets is provided. Also the follow up whether stellar
concentrations are newly detected clusters or similar work may be
done without using mid size telescopes (e.g. Boeche et al. 2003).

\section*{Acknowledgments} We are grateful to A. Henden for
more information on the USNO B1.0
calibration. We also thank the anonymous referee for his suggestions
improving our original manuscript significantly.\\
The Digitized Sky Surveys were produced at the Space Telescope
Science Institute under U.S. Government grant NAG W-2166. The
images of these surveys are based on photographic data obtained
using the Oschin Schmidt Telescope on Palomar Mountain and the UK
Schmidt Telescope. The plates were processed into the present
compressed digital form with the permission of these institutions.
The National Geographic Society - Palomar Observatory Sky Atlas
(POSS-I) was made by the California Institute of Technology with
grants from the National Geographic Society. The Second Palomar
Observatory Sky Survey (POSS-II) was made by the California
Institute of Technology with funds from the National Science
Foundation, the National Geographic Society, the Sloan Foundation,
the Samuel Oschin Foundation, and the Eastman Kodak Corporation.
The Oschin Schmidt Telescope is operated by the California
Institute of Technology and Palomar Observatory. The UK Schmidt
Telescope was operated by the Royal Observatory Edinburgh, with
funding from the UK Science and Engineering Research Council
(later the UK Particle Physics and Astronomy Research Council),
until 1988 June, and thereafter by the Anglo-Australian
Observatory. The blue plates of the southern Sky Atlas and its
Equatorial Extension (together known as the SERC-J), as well as
the Equatorial Red (ER), and the Second Epoch [red] Survey (SES)
were all taken with the UK Schmidt. Supplemental funding for
sky-survey work at the STScI is provided by the European Southern
Observatory. \\
The Guide Star Catalogue-II is a joint project of the Space
Telescope Science Institute and the Osservatorio Astronomico di
Torino. Space Telescope Science Institute is operated by the
Association of Universities for Research in Astronomy, for the
National Aeronautics and Space Administration under contract
NAS5-26555. The participation of the Osservatorio Astronomico di
Torino is supported by the Italian Council for Research in
Astronomy. Additional support is provided by European Southern
Observatory, Space Telescope European Coordinating Facility, the
International GEMINI project and the European Space

\label{lastpage}

\end{document}